\documentclass[aps,prx,twocolumn,superscriptaddress,showpacs]{revtex4-1}

\usepackage{graphicx}
\usepackage{bm}

\begin{document}

\title{Output-input coupling in thermally fluctuating biomolecular machines}

\author{Michal Kurzynski}
\email[]{kurzphys@amu.edu.pl}
\affiliation{Faculty of Physics, A. Mickiewicz University, Umultowska 85, 61-614 Poznan, Poland}
\author{Mieczyslaw Torchala}
\affiliation{Faculty of Physics, A. Mickiewicz University, Umultowska 85, 61-614 Poznan, Poland}
\affiliation{BioInfoBank Institute, Limanowskiego 24A, 60-744 Poznan, Poland}
\author{Przemyslaw Chelminiak}
\affiliation{Faculty of Physics, A. Mickiewicz University, Umultowska 85, 61-614 Poznan, Poland}

\date{\today}

\begin{abstract}
Biological molecular machines are proteins that operate under isothermal conditions hence are referred to as free energy transducers. They can be formally considered as enzymes that simultaneously catalyze two chemical reactions: the free energy-donating reaction and the free energy-accepting one. Most if not all biologically active proteins display a slow stochastic dynamics of transitions between a variety of conformational substates composing their native state. In the steady state, this dynamics is characterized by mean first-passage times between transition substates of the catalyzed reactions. On taking advantage of the assumption that each reaction proceeds through a single pair (the gate) of conformational transition substates of the enzyme-substrates complex, analytical formulas were derived for the flux-force dependence of the both reactions, the respective stalling forces and the degree of coupling between the free energy-accepting (output) reaction flux and the free energy-donating (input) one. The theory is confronted with the results of random walk simulations on the 5-dimensional hypercube. The formal proof is given that in the case of reactions proceeding through single gates, the degree of coupling cannot exceed unity. As some experiments suggest such exceeding, looking for conditions of increasing the degree of coupling over unity challenges theory. Though no analytical formulas for models involving more transition substates are available, study simulations of random walks on several model networks indicate that the case of the degree of coupling value higher than one occurs in a natural way for scale-free tree-like networks. This supports a hypothesis that the protein conformational transition networks, like higher level biological networks: the proteome and the metabolome, have evolved in a process of self-organized criticality. 
\end{abstract}


\maketitle

\section{Introduction}

An almost common conviction that biochemical processes can be interpreted in terms of the conventional chemical kinetics is based on an assumption that internal dynamics of biomolecules is fast enough to ensure reaching the partial equilibrium state before each kinetic step \cite{Kurz06}. However, at least a decade ago it was already clear that this assumption cannot be true, as besides fast vibrations the dynamics of biomolecules comprises also slower stochastic transitions between a variety of conformational substates \cite{Frau91,Beka97,Garc97,Kigo98,Kurz98}. Research of biomolecular dynamics is being developed faster and faster. Today, even in the case of small, relatively rigid water-soluble proteins, one speaks about the 'native state ensemble' \cite{Lind05,Arai06,Mori07,Vend07,Sheh07,Wuzh08,Sene08,Lange8} rather than a single native state earlier identified with the protein tertiary structure, and for very small proteins or protein fragments trials to reconstruct the actual networks of conformational transitions are realized \cite{Raoc04,Kriv04,Ryla06,Baba07,Gfel07,Noef08,Wale10}. As many as 30\% native proteins is considered to be either completely or partly intrinsically disordered \cite{Heis05,Fink05,Radi07}. Folding of such proteins takes place during the binding to targets \cite{Suga07,Wrdy09} what is essential for molecular recognition \cite{Verk02,Bonv06}. Koshland's concept of induced fit has been replaced by the 'conformational selection' concept \cite{Gohm04,Volo06,Okaz08,Lang08}. Allosteric regulation appears to have 
dynamic rather than structural nature \cite{Gohm04,Volo06,Okaz08,Lang08,Kern03,Swai06,Baha07,Cuka08,Smoc09}.

\begin{figure*}
\includegraphics[width=5.8in]{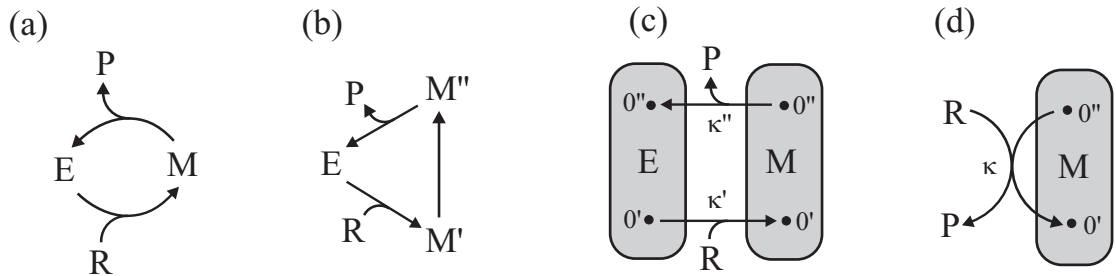}
\caption{Development of kinetic schemes of a single enzymatic reaction ${\rm R} \leftrightarrow {\rm P}$. (a)~Two-step Michaelis-Menten kinetics involving one enzyme-substrate intermediate M. (b)~Three-step Haldene's kinetics involving two intermediates. Here, M$'$ = ER and M$''$ = EP. (c)~Kinetics studied in Ref.~\cite{Kurz03} where transitions between intermediates within E and M were expanded to quasi-continuous networks of conformational transitions described by parts of Eqs.~(\ref{masteq}) and represented here by the gray boxes. The reactant and product binding-releasing reactions are assumed to be gated, i.e., they take place only in certain conformational substates represented here as black dots. (d)~Simplified kinetic scheme considered in the present paper. Two reactant and product binding-releasing reactions and a kinetics of transitions within E are replaced by a single bimolecular reaction. All reactions are reversible; the arrows indicate directions assumed to be forward (the corresponding rate constants in text are written with the subscript +).\label{fig-1}}
\end{figure*}

Because of the slow character of the conformational dynamics, both chemical and conformational transitions in an enzymatic protein have to be treated on an equal footing \cite{Kurz98} and jointly described by a system of coupled master equations
\begin{equation}\label{masteq}
    \dot{p}_l(t) = \sum_{l'}[w_{ll'}p_{l'}(t) - w_{l'l}p_{l}(t)]\,,
\end{equation}
determining time variation of the occupation probabilities $p_{l}(t)$ of the individual protein's substates (Fig.~\ref{fig-1}). In Eq.~(\ref{masteq}), $w_{l'l}$ is the transition probability per unit time from the substate $l$ to $l'$ and the dot denotes the time derivative. The conformational transition probabilities satisfy the detailed balance condition which, however, can be broken for the chemical transition probabilities controlled by concentrations of the enzyme substrates.

In the closed reactor, a possibility of a subsequent chemical transformation to proceed before the conformational equilibrium have been reached in the actual chemical state results in the presence of a transient non-exponential stage of the process and in an essential dynamical correction to the reaction rate constant describing the following exponential stage \cite{Kurz06,Frau91,Mori07,Vend07,Sheh07,Wuzh08}. In the open reactor under stationary conditions (the concentrations of reactants and products of the reaction kept constant), the general situation is more complex but for reactions gated by single conformational transition substates (Fig.~\ref{fig-1}(c)) a simple analytical theory was proposed \cite{Kurz98,Kurz03}. A consequence of the slow conformational transition dynamics is that the steady-state kinetics, like the transient stage kinetics, cannot be described in terms of the usual rate constants. This possibility was suggested almost forty years ago by Blumenfeld \cite{Blum74}. Later on, we have shown that more adequate physical quantities that should be used are the mean first-passage times between distinguished transition substates \cite{Kurz98,Kurz03}.

An application of the formalism to two coupled enzymatic reactions was considered in the context of the free energy transduction in biological molecular machines \cite{Kurz03}. We understand the word 'machine' quite generally as denoting any physical system that enables two other physical systems to perform work one on another. Under isothermal conditions, performance of work is equivalent to a transduction of free energy. Thus, molecular machines that operate under such conditions are referred to as free energy transducers \cite{Hill89}.

From a theoretical point of view, it is convenient to treat all biomolecular machines, also pumps and motors, as chemo-chemical machines \cite{Kurz06}, enzymes that simultaneously catalyze two chemical reactions: the free energy-donating reaction and the free energy-accepting one. Under isothermal conditions, all chemical reactions proceed due to thermal fluctuations: a free energy needed for their realization is borrowed from the environment and then returned to it. In fact, the biological molecular machines are Maxwell's demons: their mechanical or electrical elements are 'soft' and perform work at the expense of thermal fluctuations \cite{Vale90,Jones4,Yana07}. Of course, Maxwell's demon can operate only out of equilibrium and it is a task of the free energy-donating reaction to secure such conditions.

For the chemo-chemical machines, the degree of coupling, i.e., the ratio of the free energy-accepting (output) reaction flux to the free energy-donating (input) one was found to be also determined by the mean first-passage times between the conformational substates forming the reaction gates. As the mean first-passage times are not the quantities that could be directly determined in experiment, no experimental verification of the theory presented in Ref.~\cite{Kurz03} has been done as yet. The first goal of the present paper is to check the correctness of the theory by confronting it with results of Monte Carlo simulations performed on simple model networks of conformational transitions as well as to introduce quantities that could be determined experimentally.  

The essential motive of our studies is a trial to answer the intriguing question whether is it possible for the degree of coupling to have a value higher than unity. A dogma in the physical theory of, e.g., biological molecular motors is the assumption that for making a single step along its track the motor molecule has to hydrolyze at least one molecule of ATP \cite{Howa01}. Several years ago, this assumption has been questioned by a group of Japanese biophysicists from the Yanagida laboratory who, joining a specific nanometry technique with the microscopy fluorescence spectroscopy, shown that the myosin II head can make several steps along the actin filament per ATP molecule hydrolyzed \cite{Kita99,Kita05,Nish08}. This observation has been confirmed by some other laboratories \cite{Lipo04}, also for the dynein \cite{Koji02,Mall04,Repe06} moving along the microtubules. In Refs. \cite{Kurz03} and \cite{Kurz06}, basing on approximations carried too far, we suggested that the degree of coupling can exceed unity already for reactions proceeding through single pairs of transition substates. Here, we formally prove that it is not the case and show that the latter possibility realizes a natural way for the scale-free tree-like models of conformational transition networks with the output gate extended to more conformational substates.

\begin{figure*}
\includegraphics[width=6in]{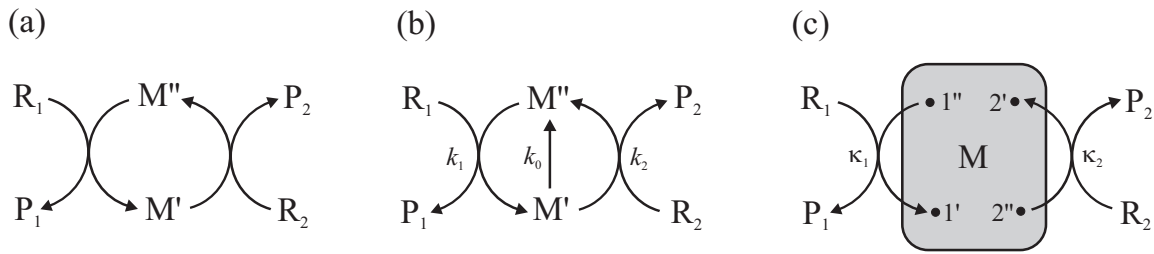}
\caption{(a)~Principle of chemochemical free energy transduction. Due to proceeding on the same enzyme, reaction ${\rm R_1} \leftrightarrow {\rm P_1}$ drives reaction ${\rm R_2} \leftrightarrow {\rm P_2}$ against its conjugate force determined by steady state concentrations of the reactant and the product. (b)~Assumption of a possible short circuit or slippage of the input vs. output reaction. (c)~Assumption of both the free energy-donating and the free energy-accepting reaction to participate in a kinetic scheme like the one shown in Fig.~\ref{fig-1}(d).\label{fig-2}}
\end{figure*}

\section{Generalized kinetic scheme of chemo-chemical machine action}

The principle of action of the chemo-chemical machine is simple \cite{Hill89}. It is a protein enzyme that catalyzes simultaneously two chemical reactions (Fig.~\ref{fig-2}(a)). Separately, each reaction takes place in the direction determined by the second law of thermodynamics, i.e., the condition that energy dissipated, determined by the product of flux and force, is positive. However, if both reactions take place simultaneously in a common cycle, they must proceed in the same direction and the direction of the first reaction can force a change of direction of the second. As a consequence, the first reaction transfers a part of its free energy recovered from dissipation performing work on the second reaction.

In formal terms, the chemo-chemical machine couples two unimolecular reactions: the free energy-donating reaction ${\rm R_1} \leftrightarrow {\rm P_1}$ and the free energy-accepting reaction  ${\rm R_2} \leftrightarrow {\rm P_2}$. Bimolecular reactions can be considered as effective unimolecular reactions on assuming a constant concentration of one of the reagents, e.g. ADP in the case of ATP hydrolysis. The input and output fluxes $J_i$ ($i$ = 1 and 2,
respectively) and the conjugate thermodynamic forces $A_i$ are defined
as \cite{Hill89}
\begin{equation}
J_i = \frac{d[{\rm P}_i]/dt}{[{\rm E}]_0}
\label{flx}
\end{equation}
and
\begin{equation}
\beta A_i = \ln K_i\frac{[{\rm R}_i]}{[{\rm P}_i]}\,,~~~~
K_i \equiv \frac{[{\rm P}_i]^{\rm eq}}{[{\rm R}_i]^{\rm eq}}\,.
\label{fce}
\end{equation}
Here, symbols of the chemical compounds in square brackets denote the molar concentrations in the steady state (no superscript) or in the equilibrium (the superscript eq). [E]$_0$ is the total concentration of the  enzyme and $\beta$ is proportional to the reciprocal temperature, $\beta \equiv (k_{\rm B}T)^{-1}$, where $k_{\rm B}$ is the Boltzmann constant. The thermodynamic forces measure the distance from the equilibrium at which they vanish. The flux-force dependence is one-to-one only if some constraints are put on the concentrations [R$_i$] and [P$_i$] for each $i$. There are two possibilities. Either the concentration of one species, say R$_i$, in the open reactor under consideration is kept constant:
\begin{equation}
    [{\rm R}_i] = {\rm const.} = [{\rm R_i}]^{\rm eq}
\label{rctcst}
\end{equation}
or is such the total concentration of the enzyme substrate:
\begin{equation}
    [{\rm R_i}] + [{\rm P}_i] = {\rm const.} = [{\rm R}]_{i0}\,.
\label{totcst}
\end{equation}
 
The free energy transduction is realized if the product $J_2A_2$, representing the output power, is negative. The efficiency of the machine is the ratio
\begin{equation} \label{efdef}
\eta = -J_2A_2/J_1A_1
\end{equation}
of the output power to the input power. In general, the degree of coupling
\begin{equation} \label{dcdef}
\epsilon = J_2/J_1\,,
\end{equation}
being itself a function of the forces $A_1$ and $A_2$, can be both positive and negative. 

Usually, the assumption of tight coupling between the both reactions is made (Fig.~\ref{fig-2}(a)). It states that the flux of the first reaction equals the flux of the second, $J_1=J_2$ thus $\epsilon = 1$. However, an additional reaction can take place between the two states M$'$ and M$''$ of the enzyme-substrates complex (Fig.~\ref{fig-2}(b)). The latter reaction can be considered either as a short circuit, the non-productive realization of the first reaction not driving the second reaction, or a slippage, the realization of the second reaction in the direction dictated by its conjugate force. The short circuit and the slippage are more favorable thermodynamically, so for the free energy transduction still to take place, the remaining reactions should be more favorable from the kinetic standpoint. 

The multiconformational counterpart of the scheme in Fig.~\ref{fig-2}(b) is shown in Fig.~\ref{fig-2}(c). Here, like in the scheme in Fig.~\ref{fig-1}(d), a network of conformational transitions within the enzyme-substrates complex is represented by the gray box and the assumption of gating by single pairs of conformational transition substates is made. The input and the output reaction fluxes are determined by stationary occupation probabilities of the gating transition substates. These can be calculated with the help of a technique of summing up the directional diagrams proposed by Terell L. Hill \cite{Hill89} who formalized an old idea of Gustav Kirchhoff. In Ref.~\cite{Kurz03}, we shown how various classes of directional diagrams for the networks of conformational substates are related to the mean first-passage times between distinguished substates. Hence, the fluxes are to be expressed in terms of appropriate mean first-passage times. In the next Section, we quote the most important results of    
Ref.~\cite{Kurz03} in the essentially changed notation facilitating their direct experimental verification.

\section{Singly gated reactions: Theoretical results}

For all the schemes shown in Fig.~\ref{fig-2}, the flux-force dependence for the two coupled reactions has a general functional form \cite{Kurz03}
\begin{equation}
J_i = \frac{1 - e^{-\beta (A_i - A^{\rm st}_i)}}
{J_{+i}^{-1} + J_{-i}^{-1} e^{-\beta (A_i - A^{\rm st}_i)}
+ J_{0i}^{-1}(K_i + e^{\beta A_i})^{-1}} \,.
\label{flxfce}
\end{equation}
The parameters $J_{+i}$, $J_{-i}$, $J_{0i}$ and $A^{\rm st}_i$ depend on the other force and are determined by a particular kinetic scheme. $A^{\rm st}_i$ have the meaning of stalling forces for which the fluxes $J_i$ vanish: $J_i(A_i^{\rm st}) = 0$. The dependence $J_i(A_i)$ is strictly increasing with an inflection point, determined by $J_{0i}$, and two asymptotes, $J_{+i}$ and $J_{-i}$ (compare Fig.~\ref{fig-4} further on). The asymptotic fluxes $J_{+i}$ and $J_{-i}$ display the Michaelis-Menten dependence on the substrate concentrations:
\begin{equation}\label{michme}
J_{+i}^{-1} = \frac{1}{k_{+i}} + \frac{K_{+i}}{k_{+i}[{\rm R}_i]}\,,~~~~
J_{-i}^{-1} = \frac{1}{k_{-i}} + \frac{K_{-i}}{k_{-i}[{\rm P}_i]}\,.
\end{equation}
Because of high complexity, we refrain from giving any formulas for the turnover numbers $k_{\pm i}$ and the apparent dissociation constants $K_{\pm i}$.

Without the loss of generality we assume that both $J_1$ and $A_1$ are positive. However, $J_2$ can be either positive or negative. Accordingly, the stalling force $A_2^{\rm st}$ should be negative or positive. Within the range of free energy transduction, $A_2^{\rm st} \leq A_2 \leq 0$ or $0 \leq A_2 \leq A_2^{\rm st}$, the flux-force dependence $J_2(A_2)$ can be convex, concave or involving an inflection point as well. The linear Onsager approximation is in general not applicable. 

The degree of coupling dependence on the forces $A_1$ and $A_2$ has a functional form
\begin{equation}
\epsilon =
\frac{1 - e^{-\beta (A_1+A_2)} + W_1(A_1)(1 - e^{-\beta A_2})}
{1 - e^{-\beta (A_1+A_2)}+ W_2(A_2)(1 - e^{-\beta A_1})}
\label{degcp}
\end{equation}
and the stalling force
\begin{equation}
    \beta A^{\rm st}_2 = \ln \frac{e^{-\beta A_1} + W_1(A_1)}{1 + W_1(A_1)}\,.
\label{stfce}
\end{equation}
The expression for $A^{\rm st}_1$ is to be obtained after replacing the index 1 with the index 2 and vice versa. The quantities $W_i(A_i)$ are measures of the slippage. The lower slippage the lower value of the corresponding $W_i$. For no slippage we simply have $A^{\rm st}_2 = -A_1$, $A^{\rm st}_1 = -A_2$, and $\epsilon = 1$ (the both fluxes $J_1$ and $J_2$ are driven by the same resultant force $A_1 + A_2$). For finite slippage, $A^{\rm st}_i$ reach the maximum, negative or positive values in the asymptotic limits of sufficiently high values of the other forces (compare Fig.~\ref{fig-6} further on).

For the simple scheme shown in Fig.~\ref{fig-2}(b), a direct calculation results in
\begin{equation}
    W_1(A_1) = k_{0+}\tau_1(A_1)\,,~~~~W_2(A_2) = k_{0-}\tau_2(A_2)\,,
\label{slis}
\end{equation}
whereas for the scheme in Fig.~\ref{fig-2}(c), the summation over diagrams  gives \cite{Kurz03}
\begin{widetext}
\begin{equation}
W_1(A_1) =  
\frac{\tau_{\rm M}(1''\!\!\leftrightarrow\!\!\{1'\!,2'\}) +
\tau_{\rm M}(1'\!\!\leftrightarrow\!\!\{1''\!,2''\})e^{-\beta A_1} + \tau_1(A_1)}
{\tau_{\rm M}(1'\!\leftrightarrow\!\{1'',2'\}) -
\tau_{\rm M}(1'\!\leftrightarrow\!\{1'',2''\})}
\label{slip}
\end{equation}
\end{widetext}
and the adequate for $W_2(A_2)$ after replacing 1 with 2 and vice versa (note the $180^{\circ}$ rotational symmetry of the kinetic scheme in Fig.~\ref{fig-2}(c)). The quantities $\tau_i(A_i)$ in Eqs.~(\ref{slis}) and (\ref{slip}) are mean transition times in the forward direction outside the enzyme-substrates complex M. For the outside transition of the form shown in Fig.~\ref{fig-1}(c), this time is given by the expression \cite{Kurz03}
\begin{eqnarray}\label{timec}
    \tau (A) &=& (k''_+)^{-1} \\\nonumber
&+& \left[ (k'_-)^{-1} + P^{\rm eq}({\rm M})
\frac{[{\rm R}]^{\rm eq}}{[{\rm R}]} \tau_{\rm E}(0'\!\leftrightarrow\! 0'')
\right] e^{-\beta A}
\end{eqnarray}
($\tau_{\rm E}(0'\!\leftrightarrow\! 0'')$ is the sum of the forward and the reverse mean first-passage times between the distinguished substates in E).
For the outside transition of the form shown in Fig.~\ref{fig-1}(d), it is simplified to
\begin{equation}\label{timed}
    \tau (A) = (k_+[{\rm R}])^{-1}\;.
\end{equation}
Above, we introduced the transition state theory rate constants defined as
\begin{eqnarray}\label{tstr}
    k''_+ &=& \kappa ''_+ p^{\rm eq}_{0''}({\rm M})/P^{\rm eq}({\rm M})\,,
\\
    k'_- &=& \kappa '_- p^{\rm eq}_{0'}({\rm M})/P^{\rm eq}({\rm M})\,,
\nonumber\\
    k_+ &=& \kappa_+ p^{\rm eq}_{0''}({\rm M})/P^{\rm eq}({\rm M})\,.
\nonumber
\end{eqnarray}
$p^{\rm eq}_l({\rm M})$ denotes the equilibrium occupation probability of the transition conformational substate $l$ in M and $P^{\rm eq}({\rm M})$ denotes the equilibrium occupation probability of the whole enzyme-substrate complex M. Notation of the transition probabilities between reaction transition substates per unit time $\kappa ''$, $\kappa '$ and $\kappa$ is explained in Fig.~\ref{fig-1}.

The quantities $\tau_{\rm M}$ in Eq.~(\ref{slip}) denote the sums 
\begin{equation}
\tau_{\rm M}(l_0\!\leftrightarrow\! \{l,l'\}) =  
\tau_{\rm M}(l_0 \!\to\! \{l,l'\}) + \tau_{\rm M}(\{l,l'\} \!\to\! l_0)   
\end{equation}
of the forward and reverse mean first-passage times that occur in the summation formula for the mean first-passage time from $l_0$ to $l$ in the network symbolized by the gray box M:
\begin{equation}\label{sumgen}
\tau_{\rm M}(l_0\!\to\! l)=
\tau_{\rm M}(\{l_0,l'\} \!\to\! l) + \tau_{\rm M}(l_0\!\to\! \{l, l'\}) 
\end{equation}
for arbitrary $l'$. Eq.~(\ref{sumgen}) is a generalization of the obvious summation formula for one-dimensional diffusion:
\begin{equation}
\tau (l_0\!\to\! l)=
\tau (l_0\!\to\! l') + \tau (l'\!\to\! l) \/,
\end{equation}
$l'$ lying between $l_0$ and $l$. The quantity $\tau_{\bf M}(l_0\!\to\! \{l, l'\})$ has a direct meaning of the mean first-passage time from $l_0$ to $l$ {\em or} $l'$. The interpretation of $\tau_{\bf M}(\{l_0,l'\} \!\to\! l)$ is more troublesome but it can be always treated as a completion of $\tau_{\bf M}(l_0\!\to\! \{l, l'\})$ to $\tau_{\bf M}(l_0\!\to\! l)$. On doing it we find two alternative relations
\begin{eqnarray}\label{interp}
\tau_{\rm M}(l_0\!\leftrightarrow\! \{l,l'\}) &=&  
\tau_{\rm M}(l_0 \!\to\! \{l,l'\}) \\
&-& \tau_{\rm M}(l \!\to\! \{l_0,l'\}) + \tau_{\rm M}(l \!\to\! l_0)\nonumber\\
&=& \tau_{\rm M}(l_0 \!\to\! \{l',l\}) \nonumber\\
&-& \tau_{\rm M}(l' \!\to\! \{l_0,l\}) + \tau_{\rm M}(l' \!\to\! l_0)\;. \nonumber
\end{eqnarray}

\section{Singly gated reactions: Comparison with Monte Carlo simulations}

The quantities $\tau_{\rm M}(l_0\!\leftrightarrow\! \{l,l'\})$ occurring in Eq.~(\ref{slip}) for $W_1(A_1)$ and the adequate for $W_2(A_2)$ can be considered as six independent parameters of the theory to be fitted in future experiments. However, the mean first-passage times occurring in Eqs.~(\ref{interp}) are not the quantities that could be directly determined experimentally. The choice of an appropriate network that models the interior of the gray box in Fig.~\ref{fig-2}(c) and positions of the input and output gates is a question of the statement of a more or less reasonable hypothesis. The simplest was stated seventy years ago by Kramers who assumed that the slowly varying intramolecular substates lie along a one-dimensional 'reaction coordinate' \cite{Frau91,Hang90}. This way of the reasoning was continued in the theory of molecular motors where the reaction coordinate was identified with the position of the motor along its track \cite{Howa01,Hill89,Astu97,Juli97}.

More complex modeling can base on statistical analyses of time series found in  molecular dynamics simulations \cite{Raoc04,Kriv04,Ryla06,Baba07,Gfel07,Noef08,Wale10} or single-molecule experiments \cite{Luxu98,Edma00,Lerc02,Flov05,Flom05,Brun05,Flom06,Kurz08}. Unfortunately, the present-day knowledge in this matter is still rather poor, so we decided first to test our theory by resorting to Monte Carlo simulations of random walks on simple, but not quite real networks. Various networks differ one from other by the geometry of links that determines an entropic contribution to the kinetics, and the variety and asymmetry of the transition probabilities $w_{ll'}$ that, following the detailed balance condition, determine an energetic contribution.   

In the beginning, for more detailed studies we chose the most regular isoenergetic network, the $n$-dimensional hypercube, the vertexes of which are labeled by sequences of the bits  $(s_1,s_2,\ldots ,s_n), ~s_i = 0,1$, and all possible transitions are related to the change of one bit and have the same probability $w$. The distance between vertexes is determined by the minimum number of edges a random walker has to pass in a walk between these vertexes. The such determined distance equals the number of the necessary bit changes. In the $n$-dimensional hypercube, there are $N = 2^n$ vertexes, each vertex has $n$ neighbors and no boundary conditions are necessary. For a reasonable time of simulations we assumed the dimension $n=5$. 

We chose the input and the output gates so as to make the free energy transduction the most effective. It takes place when moduli of both the degree of coupling (\ref{degcpz}) and the stalling force (\ref{stfce}) are maximum, i.e., the values of $W_2(0)$ and $W_1(\infty)$ are minimum. A detailed analysis indicated that it holds if the pairs of sites $1'$ and $2''$ as well as $2'$ and $1''$ lie at the closest, at the distance equal to 1. On the contrary, the diameters of the input and output gates should be larger. We chose them the largest, equal to 5 so as to lie along the diagonals of the hypercube.

\begin{figure}[b]
\includegraphics[angle=270, width=3.0in]{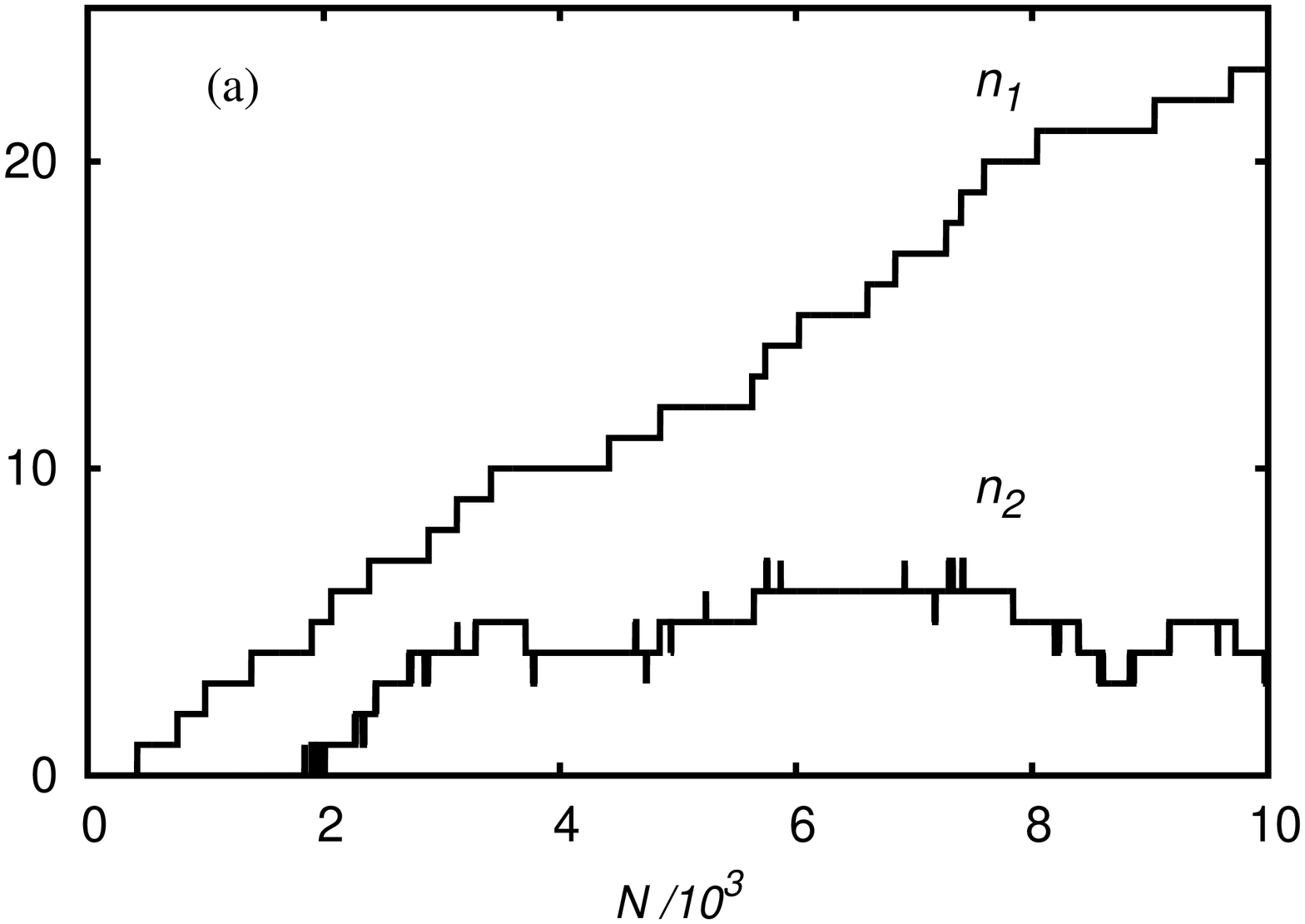}
\vspace{1pc}

\includegraphics[angle=270, width=3.0in]{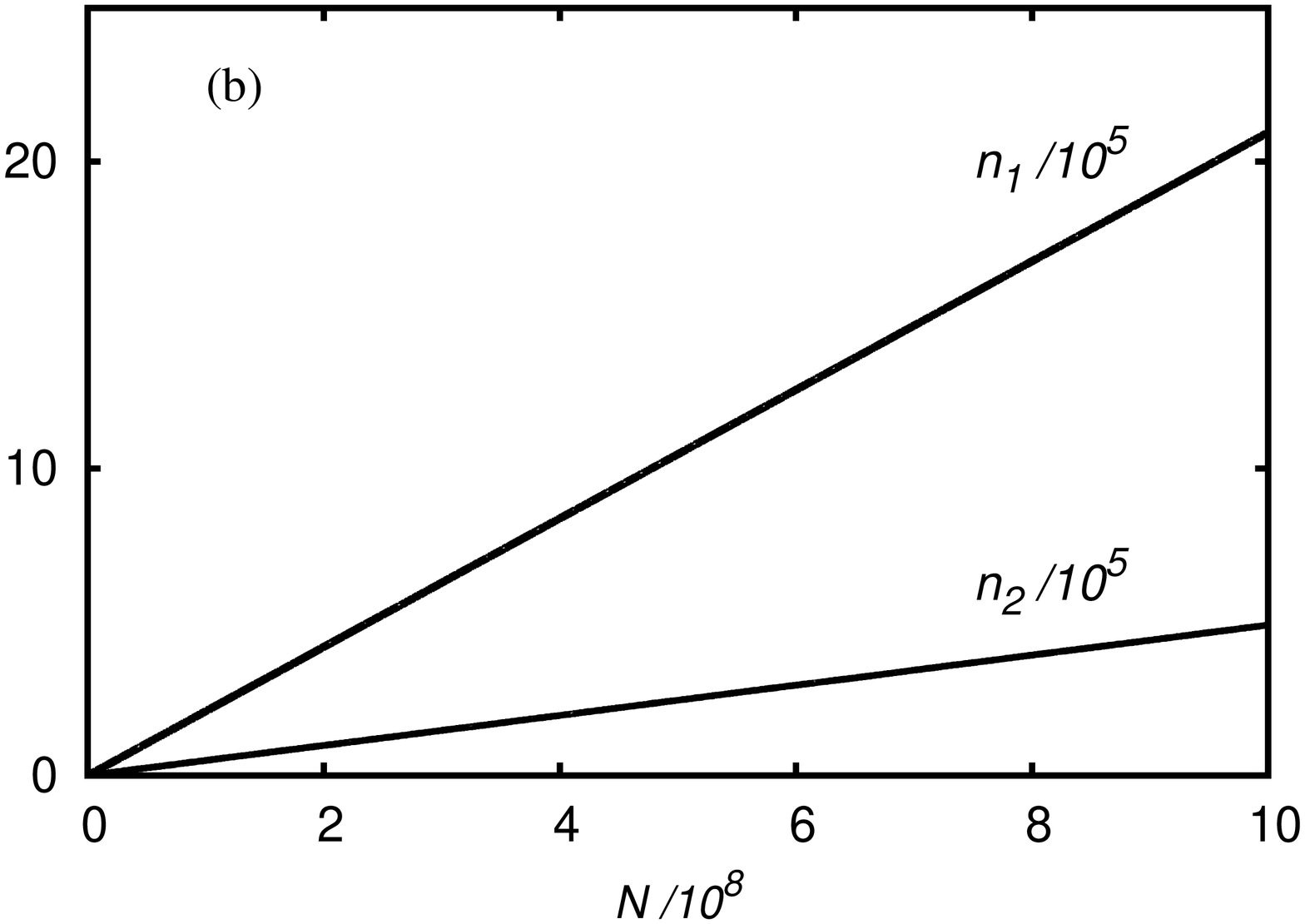}
\caption{Simulated time course of the net number of the input (${\rm R}_1 \leftrightarrow {\rm P}_1$) and the output (${\rm R}_2 \leftrightarrow {\rm P}_2$) external transitions for the 5-dimensional hypercube with gates and parameters described in text. (a) Snapshots made every step. (b) Snapshots made every $10^5$ steps.\label{fig-3}}
\end{figure}

The such determined geometry of the gates is unique and, moreover, only the values of three types of the mean first-passage times have to be known, enabling one to calculate, following Eq.~(\ref{interp}), all the quantities occurring in Eq.~(\ref{slip}) for $W_1(A_1)$ and the adequate for $W_2(A_2)$. These are the mean first-passage times of the type 
$$\tau_{\rm M}(1'' \!\to\! \{1',2'\}) = 16\,, \vspace{-1pc}$$
$$\tau_{\rm M}(1' \!\to\! \{1'',2'\}) = 80/3 \approx 26.667\,, \vspace{-1pc}$$
$$\tau_{\rm M}(1' \!\to\! 1'') = 128/3 \approx 42.667 $$
(note that for the hypercube, the mean first-passage times depend only on the distances between the initial, final and the intermediate, if any, states). All the quoted values, counted in the number of the random walk steps with the transition probability between the nearest neighbors $w = 1/n = 1/5$, were determined by simple though tedious combinatorics and checked in numerical simulations.      

\begin{figure}[t]
\includegraphics[angle=270, width=3.0in]{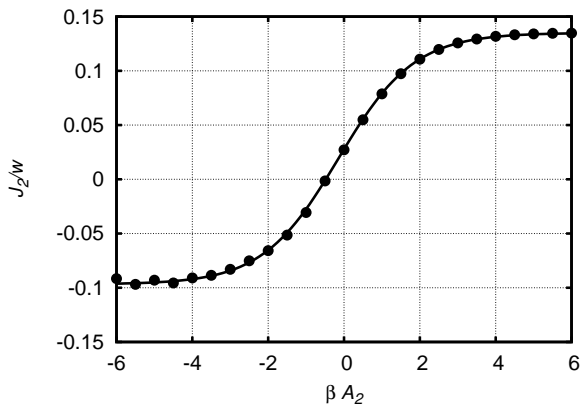}
\caption{Flux-force dependence $J_2(A_2)$ for the model and parameters described in text. The black dots denote results of the Monte Carlo simulations and the continuous line represents the fit to Eq.~(\ref{flxfce}). The free energy transduction is realized in the range $-0.48 \leq \beta A_2 \leq 0$.\label{fig-4}}
\end{figure}

We assumed the simplest, constant reciprocal forward external transition times given by Eq.~(\ref{timed}) with the use of Eq.~(\ref{rctcst}) and chose $\tau_1^{-1} = 50\,w/N$ and $\tau_2^{-1} = 30\,w/N$, counted in the reciprocal time units $w$. For $w = 1/n = 1/5$ and the equilibrium occupation probability of the each lattice vertex $1/N = 1/2^n = 1/32$, the times
$$\tau_1 = 3.200,~~~~\tau_2 = 5.333 $$ 
are one order of the magnitude shorter than the maximum mean first-passage time $\tau_{\rm M}(1' \!\to\! 1'')$ being a measure of the intramolecular relaxation time. Thus, the both reactions 1 and 2 are controlled, though not completely, by the intramolecular dynamics \cite{Kurz06}. The reciprocal reverse external transition times equal $\tau_1^{-1}$ and $\tau_2^{-1}$ multiplied by the detailed equilibrium condition-breaking exponents $\exp (-\beta A_1)$ and $\exp (-\beta A_2)$, respectively. We chose $\beta A_1 = 10$ which is a physically reasonable condition of the free energy donating reaction 1 to proceed sufficiently far from the equilibrium. 

The reciprocal external transition times multiplied by the equilibrium occupation probability $1/N$ of the transition gates determine the external transition rates. In actual simulations, to preserve equal probabilities of the forward and the reverse internal transitions in the presence of additional external transitions, $w$ had to be chosen much lower than $1/n$ and a high probability of waiting at each but one vertex had to be added.  

\begin{figure}[t]
\includegraphics[angle=270, width=3.0in]{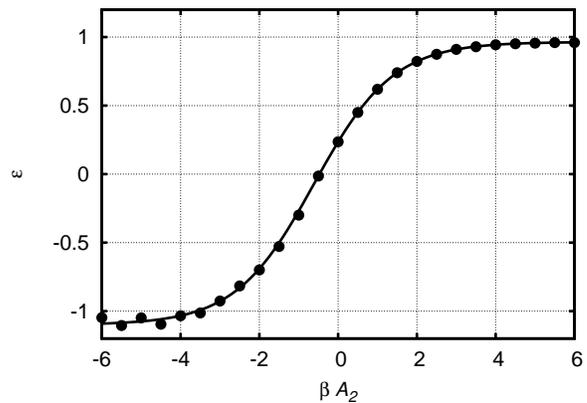}
\caption{Dependence of the degree of coupling $\epsilon$ on the output force $A_2$ for the model and parameters described in text. The black dots denote results of the Monte Carlo simulations and the continuous line is calculated following Eqs.~(\ref{degcp}), (\ref{slip}) and the adequate for $W_2(A_2)$. No fit has been performed between experiment and the theory as the latter has no free parameters. The present figure is similar to Fig.~\ref{fig-4} as for the very high value of the input force $\beta A_1 = 10$ the input flux $J_1$ remains almost constant.\label{fig-5}}
\end{figure}

\begin{figure}[b]
\includegraphics[angle=270,width=3.0in]{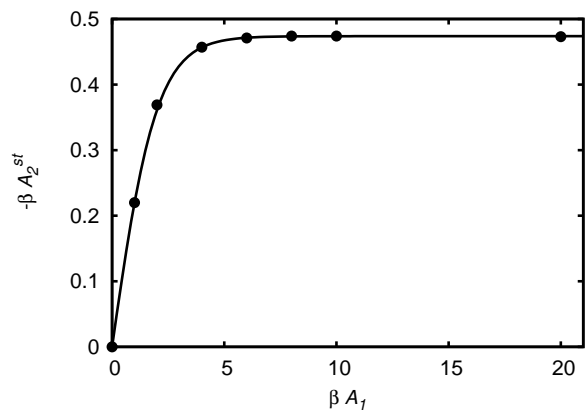}
\caption{Dependence of the stalling force $A_2^{\rm st}$ on the input force $A_1$ for the model and parameters described in text. The black dots denote results of the Monte Carlo simulations and the continuous line is calculated following Eqs.~(\ref{stfce}) and (\ref{slip}).\label{fig-6}}
\end{figure}

A typical result of a simulation of the time course of the net number of external transitions through the input and the output gates is shown in Fig.~\ref{fig-3}(a). It is clearly seen that even for such the small lattice studied, consisting of 32 vertexes, large fluctuations make determination of the input and the output fluxes in $10^4$ iteration steps impossible. Only the increase of the number of the iteration steps to $10^9$ (figure~\ref{fig-3}(b)) enables one to determine the fluxes with the error lower than 0.3\%. Dots in Fig.~\ref{fig-4} present the such determined values of the output flux $J_2$ as a function of the conjugate force $A_2$. This figure also shows that the results of our Monte Carlo experiment fit very well the theoretical prediction, Eq.~(\ref{flxfce}).

The theory presented in the previous section, in particular its main Eqs.~(\ref{degcp}), (\ref{stfce}) and (\ref{slip}), does not use any approximation and is exact. For simple networks of conformational transitions like the one considered here, we know values of appropriate mean first-passage times and can compute directly the slippage functions $W_i(A_i)$, thus the force dependences of the degree of coupling (\ref{degcp}) and the stalling force (\ref{stfce}). Figs. \ref{fig-5} and \ref{fig-6} show confrontations of the such obtained dependences with results of the Monte Carlo simulations that in the present context can be treated as experimental data. On taking into account that in Figs. \ref{fig-5} and \ref{fig-6} no fitting procedures were applied, the agreement is excellent, but it only points to the correctness of conditions at which our Monte Carlo simulations were performed. Being sure of such correctness, we can use similar simulations for testing the quality of various approximations we have to apply for actual networks. In particular, in two last sections we study by such methods effects of extending the gates to include many conformational transition substates.

\section{For singly gated reactions the degree of coupling is lower than unity}

It is not simply to evaluate the range of values that the degree of coupling (\ref{degcp}) assumes within the range of $\beta A_2$ corresponding to the free energy transduction. In Refs.~\cite{Kurz03} and \cite{Kurz06}, basing on inaccurate evaluations of the denominators in the expressions for $W_1(A_1)$ and $W_2(A_2)$, we suggested that in the case of singly gated reactions, there are possibilities of the degree of coupling modulus having a value higher than unity. Here, we present a formal proof that this modulus cannot exceed unity. The proof makes use of two relations.

From the symmetry basing the two equalities (\ref{interp}), the first relation  follows:
\begin{eqnarray}\label{rela}
\tau_{\rm M}(l_0\!\leftrightarrow\! \{l, l'\}) &-&
\tau_{\rm M}(l_0\!\leftrightarrow\! \{l, l''\}) \\
&=& \tau_{\rm M}(l\!\leftrightarrow\! \{l_0, l''\}) -
\tau_{\rm M}(l\!\leftrightarrow\! \{l_0, l'\}) \,. \nonumber
\end{eqnarray}
It secures the expression under the logarithm in Eq.~(\ref{stfce}) and the adequate for $\beta A_1^{\rm st}$ to be positive irrespectively of the sign of $W_i$. The second relation, not equivalent to (\ref{rela}), is to be derived from an equation analogous to (\ref{sumgen}) involving two intermediate nodes:
\begin{eqnarray}\label{relb}
\tau_{\rm M}(l_0\!\leftrightarrow\! \{l, l'\}) &-&
\tau_{\rm M}(l_0\!\leftrightarrow\! \{l, l''\}) \\
&=& \tau_{\rm M}(l'\!\leftrightarrow\! \{l'', l_0\}) -
\tau_{\rm M}(l'\!\leftrightarrow\! \{l'' l\}) \,. \nonumber
\end{eqnarray}
From this relation, it follows that the denominators in $W_1(A_1)$ and $W_2(A_2)$ equal each other. A consequence is that both $W_i(A_i)$ are always of the same sign, either positive or negative. 

The highest value of the degree of coupling modulus in the free energy transduction region is for $\beta A_2 = 0$. Then, Eq.~(\ref{degcp}) is simplified to
\begin{equation}\label{degcpz}
 \epsilon (0) = \frac{1}{1+W_2(0)}   
\end{equation}
independently of $\beta A_1$, and
\begin{equation} 
\beta A_1^{\rm st}(0) = 0\,.
\end{equation}
If the both $W_i$ are positive than the stalling force $\beta A^{\rm st}_2$ given by Eq.~(\ref{stfce}) is negative and the free energy transduction takes place in the region $\beta A_2^{\rm st} \leq \beta A_2 \leq 0$. The highest value of the degree of coupling (\ref{degcpz}) is positive but always lower than unity.

One could expect higher value of the degree of coupling modulus if the both $W_i$ were negative, i.e., if the denominator in (\ref{slip}) was negative, which corresponded to interchange $2'$ with $2''$ thus $J_2$ with $-J_2$. However, it is not the case. The condition of the expression (\ref{degcpz}) to be lower than minus unity is
\begin{equation}
    W_2(0)^{-1} < -1/2 \,.
\end{equation}
On substituting the explicit expression for $W_2(0)$ and taking advantage of the relation (\ref{rela}) this inequality can be rewritten as
\begin{equation}
\tau_{\rm M}(2'\!\leftrightarrow\!\{1',2''\}) +
\tau_{\rm M}(2''\!\leftrightarrow\!\{1'',2'\}) + \tau_2(0) < 0 \,.
\end{equation} 
Of course, neither $\tau_{\rm M}$ nor $\tau_2$ can be negative, so the modulus of the coupling ratio can never be higher than unity. In consequence, if the both $W_i$ are negative, the stalling force $\beta A^{\rm st}_2$ given by (\ref{stfce}) is positive and the free energy transduction also takes place, now in the region $0 \leq \beta A_2 \leq \beta A_2^{\rm st}$, however with the negative coupling ratio of the modulus also lower than 1.

\section{The degree of coupling can be higher than unity. Case of several succeeding output gates}

\begin{figure}[b]
\includegraphics[width=3.2in]{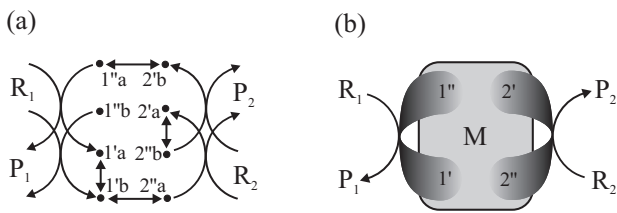}
\caption{(a) Extension of the kinetic scheme in Fig.~\ref{fig-2}(a) to two output gates. Obligatory transitions are drawn by arrows. If no other transitions are realized, the degree of coupling equals two. Otherwise, it is lower than two but possibly higher than one. (b)~Generalization of the scheme in Fig.~\ref{fig-2}(c) to a quasi-continuum of gates. The rate of the external transition within each gate can be different so that such kinds of models are referred to as the ones with 'fluctuating barriers' \cite{Kurz98}.\label{fig-7}}
\end{figure}

We proved the theorem that the value of the degree of coupling should be lower or at the most equal to unity, but only in the case when the input and output reactions proceed through single pairs of conformational transition substates.  It is reasonable to suppose that a possibility of higher degree of coupling is realized if the output gate is extended to two or more pairs of the transition substates. Indeed, in Fig.~\ref{fig-7}(a) a scheme is shown with one input gate $(1''a, 1'a)$ and two succeeding output gates $(2''a, 2'a)$ and $(2''b, 2'b)$ closed in the common cycle. It is obvious that the degree of coupling for such scheme is $\epsilon = 2$. Similar reasoning has been proposed in order to explain multiple stepping of the myosin molecule along the actin filament \cite{Kita05}. One can imagine an incorporation of a system of additional transitions and an increase of the number of the transition substates what was for the first time considered by Terada and coworkers \cite{Tera02}.

Unfortunately, even in the case of only two output gates the analytical formulas are so complex and not transparent that serious approximations are needed to be made from the very beginning. Being not able to formulate presently such approximations, we decided to apply computer experiment for a preliminary study of the problem. We performed Monte Carlo simulations starting from the 5-dimensional hypercube. For the most optimal geometry of one input and two output gates we obtained the degree of coupling $\epsilon (0)$ not higher than  0.362. Also simulations on the 9-dimensional hypercube with 512 states were not successful. We suppose that the steady state fluxes on the isoenergetic hypercube of arbitrary dimension and with arbitrary number of gates are always equivalent to steady state fluxes on some non-isoenergetic (weighted) network with various probabilities of substate occupations and transitions with the single input and output gates. Such networks are described by the theory given above and all the discussion already performed applies to them.

\begin{figure}[t]
\vspace{1pc}

\centering
\includegraphics[width=1.85in]{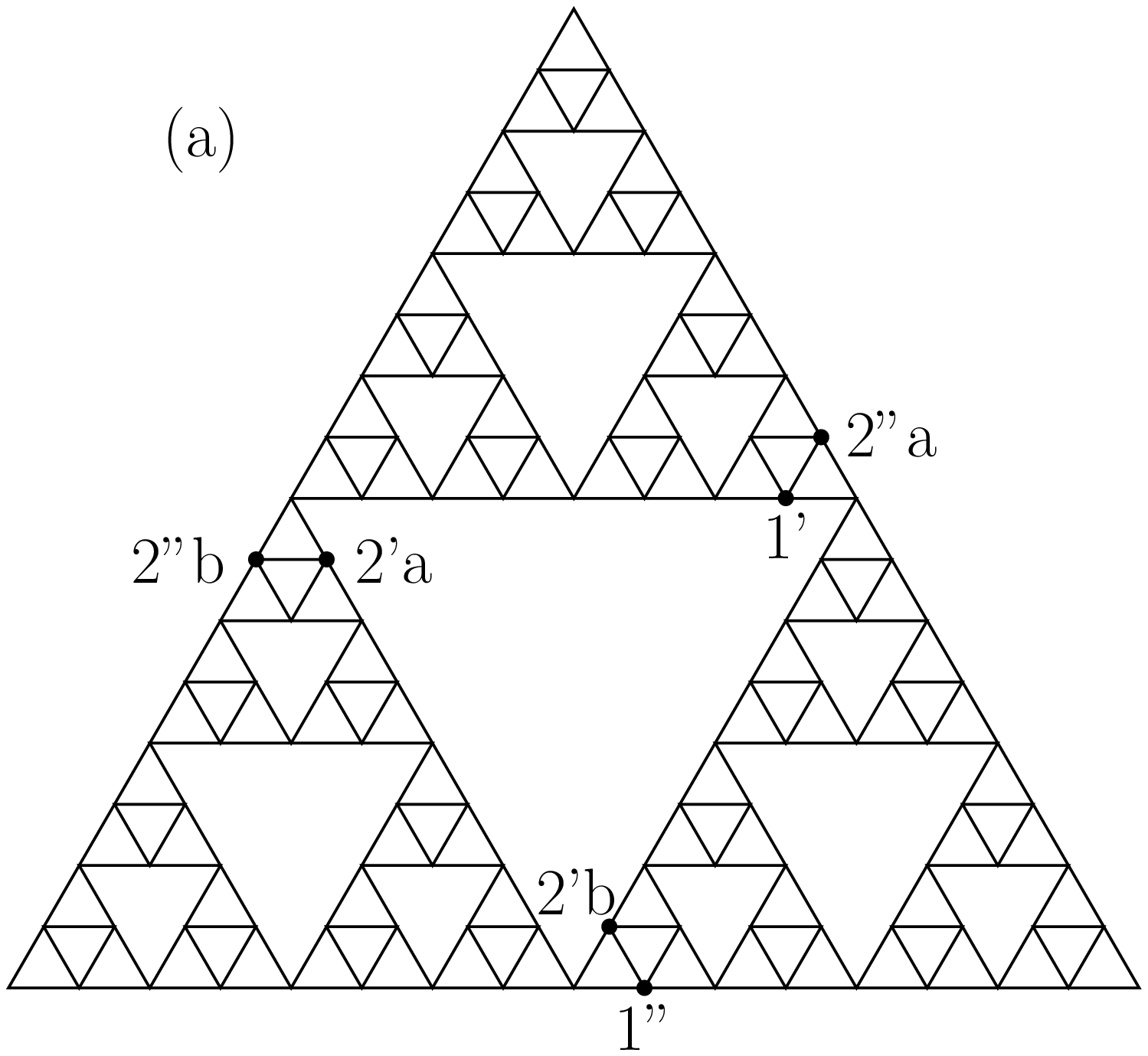}
\vspace{1pc}

\includegraphics[width=1.85in]{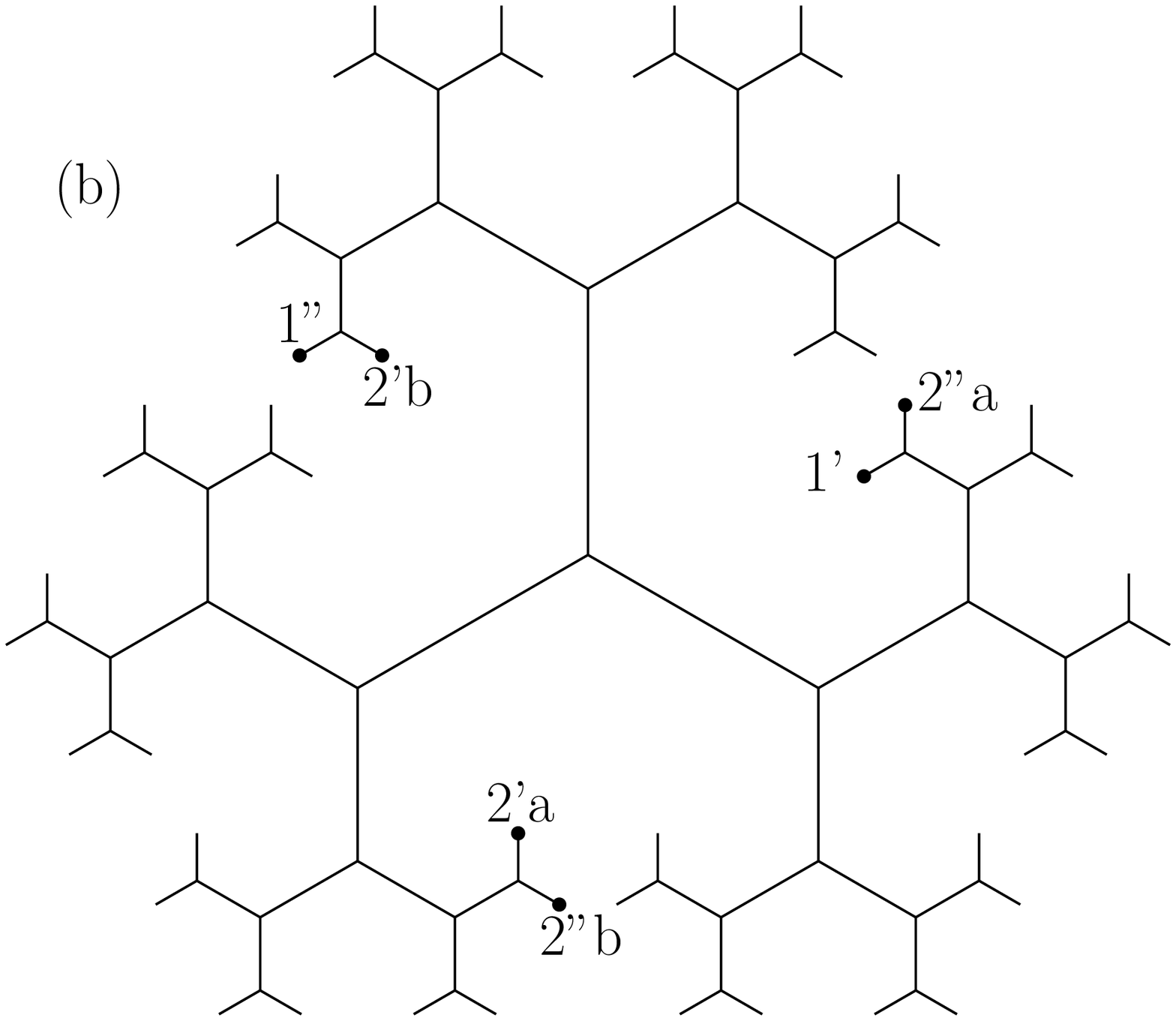}
\caption{(a)~Sierpinski gasket of the fourth order. Each vertex has four neighbors. The assumed boundary conditions allow additional transitions between the three vertexes of the largest triangle. (b)~Bethe lattice with the five shells. Each vertex besides the most external ones has three neighbors. For the most external vertexes, reflexive boundary conditions are assumed. In both networks, the geometry of the distinguished transition substates is shown.}
\label{fig-8}
\end{figure}

Still restricting ourselves to isoenergetic networks, we considered systems with more complex topology. We pointed our attention toward the networks with bottlenecks or dead ends, the diffusion on which displays long-time tiles \cite{Mont87}. The networks that model the actual conformational dynamics of proteins should display a hierarchy of relaxation times \cite{Frau91,Beka97,Garc97,Kigo98}. As an example of a network with a hierarchy of bottlenecks we considered the Sierpinski gasket (Fig.~\ref{fig-8}(a)) and as an example of a network with a hierarchy of dead ends, the Bethe lattice (Fig.~\ref{fig-8}(b)). We assumed the values of the external transition parameters as in the former section: $\tau_1^{-1} = 50\,w/N$, $\tau_2^{-1} = 30\,w/N$ and $\beta A_1 = 10$. For the Sierpinski gasket of the fourth order with boundary conditions and the system of gates shown in Fig.~\ref{fig-8}(a), we got $\epsilon(0) = 1.27$. For the Bethe lattice with the five shells and the system of gates shown in Fig.~\ref{fig-8}(b), we got $\epsilon(0) = 1.19$. We conclude that for protein machines with the stochastic dynamics described by an appropriate network of conformational transitions, the degree of coupling can in principle be higher than unity.    

The geometry of gates shown in Fig.~\ref{fig-8} was chosen with a bias against unfavorable short circuits or slippages and, simultaneously, long wandering between transitions through the successive gates. The goal was achieved in the evidently artificial way due to entropic obstacles and shortcuts. However, no obstacles and shortcuts could be needed if non-isoenergetic (weighted) networks were considered, with variable and appropriately chosen transition probabilities. In this way models with fluctuating barriers are to be obtained \cite{Kurz98}, symbolically presented in Fig.~\ref{fig-7}(b). Looking for simple model networks with the controllable and higher than unity degree of coupling that elucidates and, possibly, predicts the action of 'biomolecular gears' is certainly an important task both for theoreticians and experimentalists. In the following, last Section, we present one possible proposal.

\section{More natural mechanism of increase the degree of coupling is offered by  scale-free tree-like models}

\begin{figure}[b]
\includegraphics[width=1.85in]{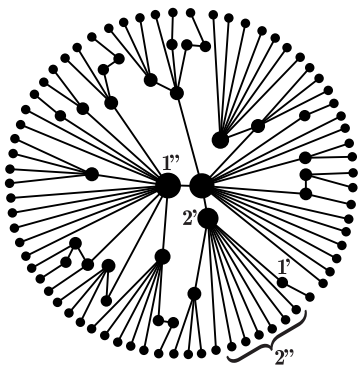}
\caption{Example of the Barab\'{a}si-Albert tree with 100 nodes and the dynamics described in text. The equilibrium occupation probabilities of the nodes are distinguished by the size of the dot and its distance from the center of the circle formed out of the highest free energy nodes with one link. The input and the output reaction transition substates are shown. Note that there are seven output transition substates $2''$.\label{fig-9}}
\end{figure}

Since the formulation by Bak and Sneppen a cellular automaton model of the Eldredge and Gould punctuated equilibrium \cite{Baks93}, the biological evolution is more and more often considered as a self-organized criticality phenomenon \cite{Bak96,Snep05}. An evolving network model of self-organized criticality was proposed by Barab\'{a}si and Albert \cite{Bara99,Alba02}. Soon it appeared that two networks of the systems biology: the proteome and the metabolome have, to a good approximation, not only the scale-free structure like the  Barab\'{a}si-Albert networks \cite{Jeon00,Jeon01} but display also a hierarchically modular, i.e., self-similar (fractal) organization \cite{Rava02,Song05}. An evolutionary mechanism has been proposed to elucidate the scale-free character of the proteome thus the metabolome \cite{Eise03}.

There are premises that also the conformational transition networks in proteins are both scale-free and hierarchically modular. The former feature is suggested by results of molecular dynamics simulations for small atomic clusters \cite{Doye02,Doye05} and by a specific spatial organization of proteins \cite{Kova05,Bode07}. The latter has been shown already in the pioneer papers from the Hans Frauenfelder laboratory \cite{Frau91} and confirmed in early molecular dynamics simulations for the very proteins \cite{Beka97,Garc97,Kigo98}. Thus, a hypothesis sounds reasonably, that also the protein conformational transition networks have evolved in the process of self-organized criticality. 

However, evolutionary speculations above are qualified. The evolving scale-free Barab\'{a}si-Albert networks have not fractal but rather small-world character \cite{Alba02}. And, indeed, such a character was also suggested for both the proteome \cite{Madi10} and the conformational transition network \cite{Doye02,Doye05,Kova05,Bode07}. Only recently, an apparent contradiction between fractality and small-worldness have been explained by application of the renormalization group technique \cite{Roze10}. It appears that a network can be fractal in a small length-scale, simultaneously having the small-world features in the large length-scale and this is the case of the proteome and, probably, the protein conformational transition networks.     

The topological structure of the flow (of probability, metabolites, energy or information) through a network is characterized by a spatial spanning tree composed of the most conducting links not involved in cycles. It is referred to as the skeleton \cite{Gohs06} or the backbone \cite{Gall07} of the network, all the rejected links being considered as shortcuts. The skeleton of the scale-free network is also scale-free but the skeleton of the self-similar network needs not be self-similar. Here a criticality feature appears important that denotes the presence of a plateau equal to unity in the mean branching number dependence on the distance from the skeleton root. The critical skeletons can be completed to self-similar scale-free networks and such is the case of the proteome \cite{Gohs06,Kigo07}. 

We state the hypothesis that such is also the case of the protein conformational transition network and that the plateau in the mean branching number versus the distance dependence corresponds to the length-scale range where the original network displays the fractal properties. Diffusion on the fractal networks is characterized by the power-low first passage time distribution and a range of such distribution was found in Monte Carlo simulations of the random walk on large Barab\'{a}si-Albert trees \cite{Chel11}. In the corresponding length-scale range the mean branching number tends to unity, i.e., the tree is in a sense equivalent to the one-dimensional chain transmitting the probability flow the fastest the possible. The latter property can be considered optimal for biological networks what justifies choosing the Barab\'{a}si-Albert trees as the object of our further considerations.

Assuming the preferential attachment rule we can construct the Barab\'{a}si-Albert trees starting from a single node and adding one node in each construction step \cite{Alba02}. In Fig.~9, an example of such a tree is shown, obtained after 99 construction steps, thus having $N = 100$ nodes. To provide the network with a stochastic dynamics described by Eq.~(\ref{masteq}), we assume the probability of changing a node to any of its neighbors to be the same in each random walk step. Consequently, the transition probability from the node $l$ to the neighboring node $l'$ 
\[
w_{l'l} = 1/k_l \,,
\]
where $k_l$ is the number of links (the degree) of the node $l$. The network with such a dynamics cannot be isoenergetic and following the detailed balance principle the equilibrium occupation probability of the node $l$
\[
p_l^{\rm eq} = k_l/\sum_{l'}k_{l'} \,.
\] 
In the scheme presented in Fig.~9, the equilibrium occupation probabilities, i.e., the values of the corresponding free energies, are distinguished by the size of the dot and its distance from the center of the circle formed out of the highest free energy nodes with one link.

As in Sections IV and VI we choose the simplest, constant mean forward external transition times given by Eq.~(\ref{timed}) with the use of Eq.~(\ref{rctcst}) and assume their values to be much shorter than the longest mean first-passage time of the internal transitions. We choose $\tau_1 = \tau_2 = 2$, which makes the external reactions almost completely controlled by the dynamics of the enzyme-substrate complex. The exit probabilities from the input ($i=1$) and the output ($i=2$) reaction transition substates in the forward direction (doubly primed) equal the reciprocal product of the external transition time and the equilibrium occupation probability of that substates, cf. Eqs.~(\ref{tstr}):  
\[
1/\tau_i p_{i''}^{\rm eq} \,.
\]
The exit probabilities from the transition substates in the reverse direction (singly primed) are similar, but must be multiplied by the factor breaking the detailed balance symmetry, determined by the external forces: 
\[
e^{-\beta A_i}/\tau_i p_{i'}^{\rm eq}\,.
\]
As in Sections IV and VI we choose $\beta A_1 = 10$ which makes the exit probability from the transition substate $1'$ negligible. The exit probabilities from all the remaining transition states are much higher than the internal transition probabilities to the neighboring substates. For securing the sum of all transition probabilities from a given node to be one in the actual simulation step, all the discussed transition probabilities are appropriately renormalized and a high probability of waiting at each but one node is added.

From what was told above it follows that both the internal transition probabilities and the exit probability from a given transition substate are inversely proportional to its degree $k_l$. This fact implies a strategy of choosing the output gates. For the resultant output flux $J_2$ at the zero external force ($\beta A_2 = 0$) to be the highest, the forward reaction transition substate $2''$ should have the lowest degree and be realized many times whereas the reverse reaction transition substate $2'$ should have a higher degee and be realized singly. As concerns the forward reaction transition substate $1''$, it should lie close to $2'$ and, because under the pumping conditions in the steady state its occupation decreases essentially, its equilibrium occupation should be the highest, i.e., it should be the main hub. The reverse reaction transition substate $1'$ need not be highly occupied but it should lie close to any of the substates $2''$. The gates chosen in accordance with this strategy are shown in Fig.~9.

In Figs.~10 (a) and (b), typical results of Monte Carlo simulations of the net number of external transitions through the input and the output gates, respectively, are shown. It is worth pointing attention to the 'devil's staircase' form of the input flux in the range of medium transition times, what is characteristic for diffusion on fractal networks. The absence of longer transition times results from the small-worldness effects discussed above as well as the boundary conditions. The absence of shorter transition times results from the finite distance between the starting and the ending node as well as admitting additional external transitions through the output gates.

\begin{figure}
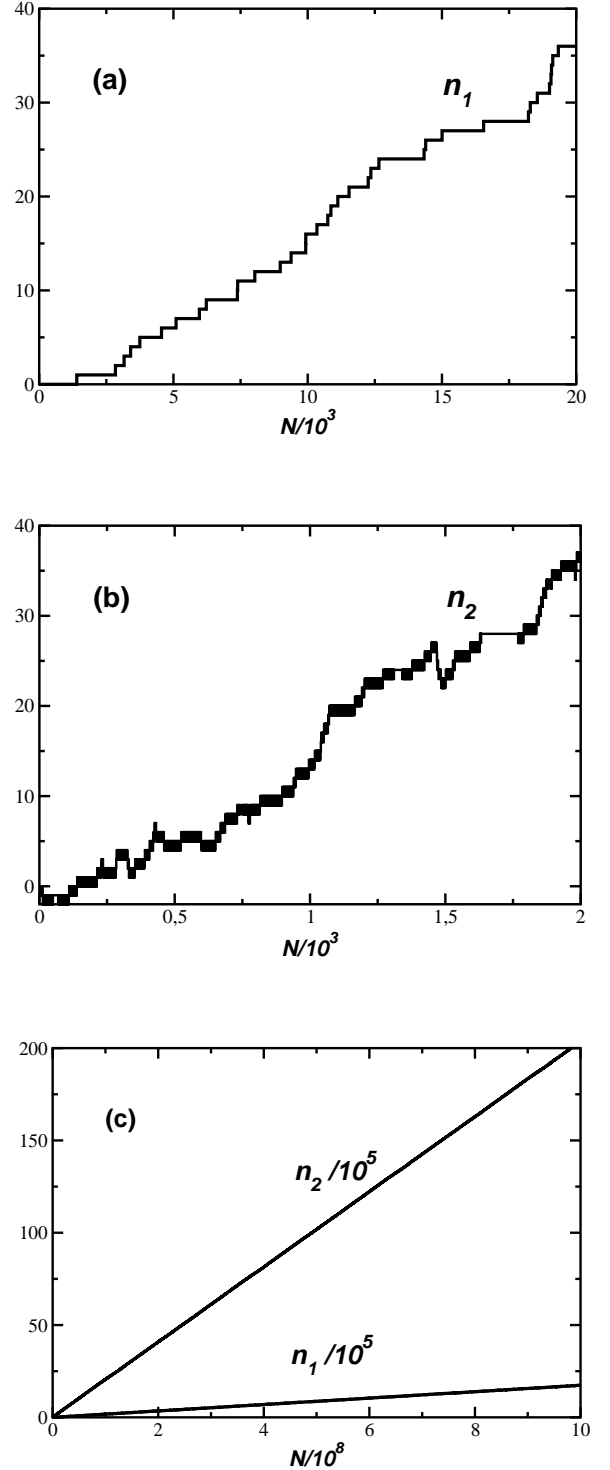


\vspace*{2pc}
\includegraphics[width=3in]{fig10a}
\vspace{2.5pc}

\includegraphics[width=3in]{fig10b}
\vspace{2.5pc}

\includegraphics[width=3in]{fig10c}
\caption{Simulated time course of the net number of the input and the output external transitions for the model presented in Fig.~\ref{fig-9} and parameters described in text ($\beta A_2 = 0$). (a)~Snapshots made every step for the input flux $n_1$. (b)~Snapshots made every step for the output flux $n_2$. Many succeeding forward and reverse transitions are seen as broadened lines. (c)~Snapshots for $n_1$ and $n_2$ made every $10^5$ steps.\label{fig-10}}
\end{figure}

In the output flux, numerous reverse transitions are observed. Nevertheless, there are more forward transitions on the average per one input flux transition. In Fig.~10~(c), both fluxes are presented in much longer time scale from which the value of the output-input degree of coupling can be easy evaluated for $\epsilon (0) = 11.0$. The main reason for such a high value of $\epsilon$ is a large representation of medium transition times in the input flux confronted with an approximately exponential distribution of transition times in the output flux.

Fig.~11 shows how the degree of coupling $\epsilon$ decreases with the output force $A_2$. For the model and the parameters assumed, the stalling force can be  evaluated for $- 1.4~k_{\rm B}T$ units. It is worth noting an excellent fit to  Eq.~(\ref{flxfce}) what, on taking into account constancy of the input flux $J_1$ within the region considered, suggests that the formula (\ref{flxfce}) remains applicable for any biomolecular machine.    

\begin{figure}
\includegraphics[width=2.85in]{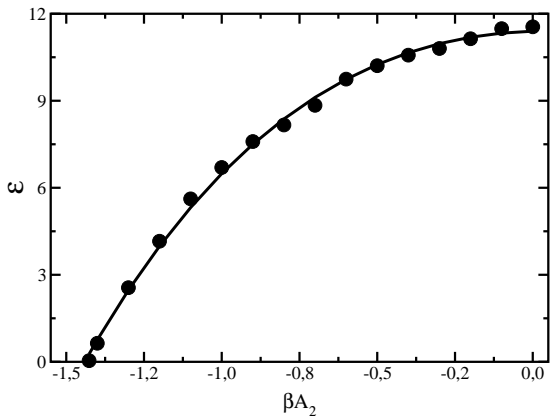}
\caption{Dependence of the degree of coupling $\epsilon$ on the output force $A_2$ for the model presented in Fig.~\ref{fig-9} and parameters described in text. The black dots denote results of the Monte Carlo simulations restricted to the free energy transduction region and the continuous line represents the fit to Eq.~(\ref{flxfce}).\label{fig-11}}
\end{figure}

\section{Concluding remarks}

Only in a few recent years, some trials have been undertaken to determine conformational transition networks in native proteins. It is why in the present paper we restricted our attention to model networks. Our goal was to calculate and simulate the degree of coupling between the free energy-accepting and the free energy-donating reaction flux in protein molecular machines. Exact theoretical formulas were possible to be obtained only for reactions proceeding through single pairs (the gates) of conformational transition substates. The theory predicts the value of the degree of coupling not exceeding unity. However, in Monte Carlo simulations on simple scale-free tree-like networks we shown that on increasing the number of the output gates one can easily obtain the degree of coupling much higher than unity. In other words it means that 'biomolecular gears' are possible.  

Nevertheless, the degree of coupling for most protein machines is lower or equal to unity. Simultaneously, most protein enzymes display the Michaelis-Menten dependence of the asymptotic fluxes on the substrate concentration. Gating the reactions by single pairs of conformational transition substates is a necessary condition for the conformationally fluctuating enzymes to obey the Michaelis-Menten kinetics \cite{Kurz98,Kurz03}. There are thus solid grounds to suppose that the theory presented in Section III is applicable in the description of action of most biological machines. Doubts can be settled by analysis of time correlation functions of the dichotomic noise observed in appropriate single-molecule experiments \cite{Luxu98,Kurz08}.   

Of course, networks with gates comprising single transition substates should be treated only as effective ones. The actual networks of conformational transitions are certainly much more complex. Various networks and systems of gates lead to the same or similar values of the quantities $\tau_{\rm M}(l_0\!\leftrightarrow\! \{l,l'\})$ in the expressions for the slippage functions $W_i(A_i)$. Similarly, various networks make identical predictions of the statistical properties of the dichotomous noise observed \cite{Brun05,Flom06}. It is a task for theoreticians to propose an algorithm of constructing the minimum effective networks that interpret the flux-force characteristics of the particular classes of protein machines. 

We tried to justify a hypothesis that the protein conformational transition networks, like higher level biological networks: the proteome and the metabolome, have evolved in a process of self-organized criticality. A proposal follows from it to adopt evolving scale-free trees for the universal models of the conformational transition networks in the biomolecular machines. We assumed that the free energy-donating reaction (usually, the ATP hydrolysis) is singly gated and proceeds through the main hub. In fact, the dependence of both the input and the output fluxes on the ATP concentration found in our simulations is of the Michaelis-Menten form, what agrees with many experiments. The universality of the ATP hydrolysis is to be confronted with the fact that the main hub is very stable and evolves slowly. On the other hand, nodes with low connectivity evolve faster and can be fitted evolutionary, being good candidates for, if need be, either single or multiple entrance gate of the free energy-accepting process.

\begin{acknowledgments}
This study has been supported in part by the Polish Ministry of Science and Higher Education (project N~N202~180038). M.T. thanks additionally the Foundation for Polish Science for a FOCUS fellowship. 
\end{acknowledgments}

\end{document}